\documentclass[12pt]{article}

\begin{document}
\begin{center}
{\Large {\bf Multi-Lagrangians for Integrable Systems}}\\[4mm]
{\Large {\bf Y. Nutku and M. V. Pavlov}}\\[2mm] Feza G\"ursey
Institute P.O.Box 6 \c{C}engelk\"oy, Istanbul 81220 Turkey\\
May 7, 2001
\end{center}


\noindent {\bf Abstract}

We propose a general scheme to construct multiple Lagrangians for
completely integrable non-linear evolution equations that admit
multi-Hamilton\-ian structure. The recursion operator plays a
fundamental role in this construction. We use a conserved quantity
higher/lower than the Hamiltonian in the potential part of the new
Lagrangian and determine the corresponding kinetic terms by
generating the appropriate momentum map.

This leads to some remarkable new developments. We show that
nonlinear evolutionary systems that admit $N$-fold first order
local Hamiltonian structure can be cast into variational form with
$2N-1$ Lagrangians which will be local functionals of Clebsch
potentials. This number increases to $3 N -2$ when the Miura
transformation is invertible. Furthermore we construct a new
Lagrangian for polytropic gas dynamics in $1+1$ dimensions which
is a {\it local} functional of the physical field variables,
namely density and velocity, thus dispensing with the necessity of
introducing Clebsch potentials entirely. This is a consequence of
bi-Hamiltonian structure with a compatible pair of first and third
order Hamiltonian operators derived from Sheftel's recursion
operator.

\section{Introduction}

In this paper we shall point out a general technique for the
construction of inequivalent solutions to the inverse problem in
the calculus of variations. We shall show that completely
integrable partial differential equations in $1+1$ dimensions that
admit multi-Hamiltonian structure can be cast into variational
form with multiple Lagrangians. It is remarkable that all these
new Lagrangians can be obtained directly from our present
knowledge of complete integrability of the evolutionary system
without doing any new calculations!

One of the important properties we expect from a completely
integrable system is multi-Hamiltonian structure. A vector
evolutionary system can then be cast into Hamiltonian form in more
than one way
\begin{equation}
u^i_{t_{\aleph+\alpha-1}} = \{u^i, H_\alpha \}_\aleph =
 J^{ik}_{\aleph}  \delta_k H_\alpha \qquad
\left\{ \begin{array}{l} i=1,2,...,n \\ \aleph=1,2,... N. \\
\alpha=-1,0,1,..., \infty
\end{array} \right. \label{hameq}
\end{equation}
where the variational derivative is denoted by $\delta_k\equiv
\delta /\delta u^k$ and $J$ is a matrix of differential operators
satisfying the properties of a Poisson tensor, namely
skew-symmetry and Jacobi identity. For integrable systems there
exists more than one such Hamiltonian operator
and Hamiltonian function as the respective Hebrew and Greek
indices indicate. Then, by the theorem of Magri \cite{magri}
completely integrable systems admit infinitely many conserved
Hamiltonian functions which are in involution with respect to Poisson
brackets defined by compatible Hamiltonian operators.

The essential element in the multi-Hamiltonian approach to
integrability is the construction of the Hamiltonian operators
themselves. Fortunately this is a rich subject \cite{dorfman} that
can be put to good use. We shall be interested in the consequences
of multi-Hamiltonian structure on the Lagrangian formulation of
completely integrable evolutionary equations. We shall work in the
opposite direction to the traditional approach of deriving
Hamiltonian structure from a Lagrangian. The crucial fact that we
shall exploit is the relationship between Hamiltonian operators
and Dirac brackets \cite{dirac} for degenerate Lagrangian systems
which was first pointed out by Macfarlane \cite{mac}. In the case
of completely integrable systems we have much more information on
Hamiltonian structure than Lagrangian and it became clear only
recently \cite{pavlov}, \cite{nhepth}, \cite{pavlov2} how we can
construct multiple Lagrangians for systems that admit
multi-Hamiltonian structure. We shall now present the general and
most simple technique for generating these new Lagrangians.

\section{Multi-Lagrangians}
\label{sec-main}

Evolutionary systems (\ref{hameq}) cannot be cast into variational
form with a local expression for the Lagrangian using the velocity
fields $u^i$ alone but require the introduction of Clebsch
potentials. In $1+1$-dimensions the general expression for Clebsch
potentials is given by
\begin{equation}
u^i =  \phi^i_x
\label{clebsch}
\end{equation}
and in this paper we shall only consider Lagrangians that are
local functionals of these potentials. In the time-honored way we
shall split the Lagrangian density for eqs.(\ref{hameq}) into two
\begin{equation}
{\cal L} =  {\cal T} - {\cal V} \label{lagtot}
\end{equation}
that consist of the kinetic and potential pieces respectively. For
the first Lagrangian density, an enumeration which will become
clear presently, the kinetic term is always given by
\begin{equation}
{\cal T}_{1} = g_{ik} \;  \phi^i_t \, \phi^k_x \label{kinetic}
\end{equation}
where $g_{ik}$ are constants with $\det g_{ik} \ne 0$ and
\begin{equation}
{\cal V}_{1} = 2 {\cal H}_1 \label{potential}
\end{equation}
is the Hamiltonian density. We note that the Hamiltonian function
that appears in (\ref{hameq}) is the space integral of the
density. We shall number the conserved Hamiltonians by reserving
the subscript $1$ to the ``usual" Hamiltonian function but of
course there exists conserved quantities such as Casimirs and the
momentum which are of lower order. In fact, for complete
integrability, an $n$-component vector evolutionary system
(\ref{hameq}) must admit $n$ infinite series of conserved
Hamiltonians. We shall denote their densities by
\begin{equation}
{\cal H}_{\alpha ; [i]} \qquad \qquad i=1,2,..n; \quad
\alpha=-1,...,\infty  \label{series}
\end{equation}
and recall that each series starts with a Casimir
\begin{equation}
 {\cal H}_{-1 ; [i]} = g_{ik} u^k   \label{casimir}
\end{equation}
which will carry the label minus one. One of these series is
distinguished in that it contains the ``usual" Hamiltonian
function which is the one that appears in eq.(\ref{hameq}). For
the $2$-component systems that we shall discuss in this paper
these are the Eulerian and Lagrangian series. We note also that
the two series may coincide up to a relabelling dictated by the
recursion operator. This is in fact the case for the $\gamma=2$
case of gas dynamics and in most examples of completely integrable
dispersive equations except the Boussinesq equation.

The potential part of the Lagrangian does not depend on the
velocities and from eq.(\ref{kinetic}) it follows that the Hessian
$$ \det \left| \frac{\partial^2 {\cal L}_1}{\partial \phi^i_t \,
\partial \phi^k_t} \right| = 0 $$
vanishes identically. We have therefore a degenerate Lagrangian
system and in order to cast it into Hamiltonian form we must use
Dirac's theory of constraints \cite{dirac}, or the covariant
Witten-Zuckerman theory \cite{witten,zuck} of symplectic
structure. In particular, the first Hamiltonian operator obtained
from the first Lagrangian is given by
\begin{equation}
J^{ik}_{1}  = g^{ik} \, D   \qquad D \equiv \frac{d}{d x}
\label{j0}
\end{equation}
where $g^{ik}$ is the inverse of the coefficients in the kinetic
part of the first Lagrangian (\ref{kinetic}) which is
non-degenerate.

The construction of multiple Lagrangians relies on the use of the
Lenard recursion relation which is implicit in eqs.(\ref{hameq})
that in the Greek and Hebrew indices we have a symmetric matrix
\begin{equation}
J^{ik}_{[\aleph}  \delta_{|k|} H_{\alpha]} =0 \label{lenard}
\end{equation}
where square brackets denote complete skew-symmetrization and bars enclose
indices which are excluded in this process.
Provided we can invert these Hamiltonian operators,
we can construct recursion operators
\begin{equation}
   R_{\;\aleph_2 \; \;k}^{\aleph_1 \;\; i} = J_{\aleph_2}^{im}
   (J^{mk}_{\aleph_1})^{-1}
\label{recop}
\end{equation}
that map gradients of conserved Hamiltonians into each other (\ref{lenard}).

For the construction of Lagrangians we start with the crucial
observation that the first Lagrangian is of the form
\begin{equation}
{\cal L}_1 = {\cal H}_{-1 [i]} \, \phi^i_t - 2 {\cal H}_1
\label{observe}
\end{equation}
which is manifest from (\ref{kinetic}). The original fields that
enter into the evolutionary system (\ref{hameq}) are Casimirs
which is evident from the subscript minus one. The second
Lagrangian will be of the same general structure as
(\ref{observe}) if we further suppose that eqs.(\ref{hameq}) can
be written in bi-Hamiltonian form. Thus there will exist $H_2$
which is the next conserved Hamiltonian function in the hierarchy
and the momentum $H_0$ which comes after Casimirs. The higher
Lagrangian should simply be $$ {\cal L}_2 = {\cal H}_{0 [i]} \,
\phi^i_t - 2 {\cal H}_2$$ but there is an important refinement
that we need to insert here. It is not the conserved density but
rather the momentum map that enters into the kinetic part of the
Lagrangian. The two differ only by total derivatives which is
irrelevant in the context of conservation laws and therefore
generally skipped over. However, these divergence terms are of
crucial interest as the momentum map in the theory of symplectic
structure. We shall show that given $\alpha^{th}$ local
Hamiltonian structure, the full new Lagrangian is simply given by
\begin{equation}
{\cal L}_\alpha = \{ {\cal H}_{\alpha-2 [i]} + ({\cal G}_{\alpha
-2  [i]} )_x \} \phi^i_t - 2 {\cal H}_{\alpha} \label{genexp}
\end{equation}
where ${\cal G}_{\alpha [i]}$ is a functional of the potentials.
The coefficient of $\phi^i_t$ above is the momentum map and this
is the only calculation necessary to find the new Lagrangian.

The fact that it is the momentum map rather than the conserved
density that plays an important role in the Lagrangian can be seen
at the level of the first Lagrangian. Now the Casimirs play the
role of the momentum map and they are used to construct the next
higher conserved quantity according to the construction of the
canonical energy-momentum tensor
$$ {\cal H}_0 = \frac{\partial {\cal L}_1}{\partial \phi^i_t}
\phi^i_x = {\cal H}_{-1 [i]} u^i = \frac{1}{2} g^{ik} {\cal H}_{-1
[i]}{\cal H}_{-1 [k]} $$ which is the momentum. This classical
result for Lagrangians linear in the velocity can be generalized
at each level we have a higher Lagrangian. We have
\begin{equation}
2 {\cal H}_{\alpha-1} = g^{ik} [  {\cal H}_{\alpha-2 [i]} + ({\cal
G}_{\alpha-2 [i] } )_x  ]  {\cal H}_{-1 [k]} \label{check}
\end{equation}
ending at the level where a local Lagrangian is no longer
possible. In fact the validity of this equation is directly
related to the existence of the Lagrangian. If a check of
(\ref{check}) fails for some $\alpha$, then there exists no local
Lagrangian at $\alpha^{th}$ level.

Now we come to an important reservation that our new Lagrangians
will necessarily carry. The Euler equations that follow from the
variation of the action with the second Lagrangian will be
\begin{equation}
 R_{\;2 \; \;k}^{1 \;\; i} \,
  \left[ u^k_{t} - J^{km}_{1}  \delta_m H_1 \right]=0
 \label{crucial}
\end{equation}
so that the first variation of the second action will certainly be
an extremum for the original equations of motion (\ref{hameq}) but
the Euler equations (\ref{crucial}) require something weaker,
namely linear combinations of functionals in the kernel of the
recursion operator can be added to the right hand side of the
equations of motion and the new action will still be an extremum.

From this construction it is manifest that for every Hamiltonian
function in the infinite hierarchy of conserved Hamiltonians that
we have for completely integrable systems, there exists a
degenerate Lagrangian (\ref{genexp}) that yields the equations of
motion as its Euler equation up to functionals in the kernel of
the recursion operator. The number of Lagrangians that can be
constructed in this way is therefore infinite in number. Given
bi-Hamiltonian structure we have two local Hamiltonian operators
but the Lenard recursion operator (\ref{recop}) is non-local.
However, the special form of the first Hamiltonian operator
(\ref{j0}) leads to a local expression for the second Lagrangian
in terms of Clebsch potentials. But it is clear that the repeated
application of the recursion operator will require the
introduction of non-local terms in higher Lagrangians. Strictly
speaking, this is not a problem because the original Lagrangian is
itself non-local in terms of the velocity fields $u^i$ which are
the original variables. We swept this problem under the rug by
introducing Clebsch potentials. Higher Lagrangians for
evolutionary equations (\ref{hameq}) can be written in local form
by introducing potentials for the Clebsch potentials themselves!
If, however, the equations of motion admit $N$ local Hamiltonian
operators, then our construction guarantees the existence of $N$
Lagrangians which are local functionals of the Clebsch potentials.
Thus we have

{\bf Theorem 1} {\it A completely integrable system that admits
$N$-fold local first order Hamiltonian structure can be given $N$
different variational formulations with degenerate Lagrangians
that are local functionals of the Clebsch potentials}.

By a convenient abuse of language we claim that we have a
Lagrangian for an equation that involves fields when the
Lagrangian is in fact only a functional of the Clebsch potentials
for these fields. Then we have the audacity to put in by hand the
expression for the fields in terms of potentials after the
variation! This can be at best only a shorthand for the real
variational principle where we must impose the relationship
between the fields and their potentials through Lagrange
multipliers.

So far we have been guilty of this abuse ourselves. But now we
must say that the first Lagrangian is actually
\begin{equation}
{\cal L}_{\ge 1}^{full} = {\cal L}_1(\phi^i, \phi^i_x,
\phi^i_{xx}...) + \lambda_i ( u^i - \phi^i_x ) \label{reallag}
\end{equation}
so that upon variation with respect to all the variables $\phi^i,
u^i, \lambda_i$ we get (\ref{clebsch}), $\lambda_i=0$ and we
arrive at the equations of motion (\ref{hameq}) expressed in terms
of the original fields $u^i$ without fudging.

Now this obvious observation may seem correct but naive, however,
we shall now find that it dramatically increases the number of new
Lagrangians we can construct for integrable systems.

For every evolutionary equation that admits, say for simplicity,
bi-Hamil\-tonian structure there exists a differential
substitution
\begin{equation}
u^i=M^i(r^k, r^k_x, ...) \label{m}
\end{equation}
that brings the second Hamiltonian operator to the canonical form
(\ref{j0}) of Darboux. This differential substitution is a Miura
transformation. Strictly speaking the theorem of Darboux remains
unproved in field theory where the number of degrees of freedom is
infinite but we shall assume it. Miura transformation works in a
direction opposite to the usual action of the recursion operator.
It leads to Hamiltonian equations
\begin{equation}
r^i_{t} = \{r^i, H_0\}_1 =  g^{ik} D \, \delta_{r^k}
H_0\Big|_{u^m=M^m(r^n)} \label{hameqmod}
\end{equation}
where $H_0$ is the momentum for eqs.(\ref{hameq}) expressed
through (\ref{m}). These are modified equations, different from
the original equations, but the two sets are related by
\begin{equation}
u^i_{t} - J_2^{ik} \delta_{u^k} H_0 (u) = {\cal O}^i_j (r^l) \,
\Big[ r^j_{t} - J_1^{jk} \delta_{r^k} H_0 \Big|_{u^m=M^m(r^n)}
\Big] \label{mm3}
\end{equation}
up to functions in the kernel of some matrix differential operator
${\cal O}^i_j$.

A comparison of eqs.(\ref{crucial}) and Miura's relation
(\ref{mm3}) shows us that using the differential substitution of
Miura we can obtain new Lagrangians for nonlinear evolution
equations that admit multi-Hamiltonian structure in the opposite
direction to our earlier construction. Transforming to the
variables $r^i$ and using Clebsch potentials
\begin{equation}
r^i = \psi_x^i \label{newclebsch}
\end{equation}
we can write the classical Lagrangian for the modified system
(\ref{hameqmod})
\begin{equation}
{\cal L}_1^{modified} = g_{ik} \psi_x^i  \psi_t^k - 2 {\cal H}_0
\Big|_{u^m=M^m(\psi_x,\psi_{xx}...) } \label{lagmod}
\end{equation}
where the labelling of ${\cal H}_0$ refers to its expression in
the original variables $u^i$ but these need to be substituted for
in terms of $r^i$ according to (\ref{m}) and expressed through the
potentials (\ref{newclebsch}). We note that the Casimirs $ r^i =
\psi^i_x$ for the modified system are absent in the polynomial
${\cal H}_\alpha(u)$ hierarchy. We would expect naively that the
Euler equations resulting from the first variation of the action
with the Lagrangian (\ref{lagmod}) would result in the modified
equations (\ref{hameqmod}). This would indeed be the case if we
were to impose the constraint between the fields $r^i$ and their
potentials $\psi^i$ as in (\ref{reallag}) but now using
(\ref{newclebsch}). However, by imposing the constraint through
Miura's differential substitution
\begin{equation}
{\cal L}_0^{full} = {\cal L}_1^{modified}( \psi^i_x,
\psi^i_{xx}...) + \lambda_i [ u^i - M^i(\psi^i_x,\psi^i_{xx}...) ]
\label{reallag2}
\end{equation}
we obtain a new Lagrangian for the original equations
(\ref{hameq}) in the original variables $u^i$. We shall use this
construction to derive new Lagrangians, in particular for KdV in
section \ref{sec-kdv}. It is evident that this construction can be
extended when there exists multi-Hamiltonian structure but, as we
shall find in the example of KdV, sometimes it is possible to
arrive at local Lagrangians using non-local Hamiltonian operators
as well. Now we conclude

{\bf Theorem 2} {\it The first Lagrangian of every modified
equation obtained through a Miura transformation will serve as a
new zeroth Lagrangian for the original equations of motion
provided the constraint between the fields and their potentials is
imposed through a Miura-type differential substitution. For
$N$-fold Hamiltonian structure there exists $N-1$ such new
Lagrangians}.

Miura transformation is in general not invertible because it is a
differential substitution. But there exists interesting examples
where it reduces to a point transformation which is invertible. In
that case we can construct $N-1$ further Lagrangians.

We conclude that for an evolutionary system that admits $N$ fold
first order Hamiltonian structure, the number of different
variational principles where the first variation will be an
extremum by virtue of the original equations of motion is $2N-1$
and in the case Miura transformation is invertible $3N-2$. We
illustrate this situation for the case of bi-Hamiltonian structure
in tables \ref{table1a} and \ref{table1b}. The general situation
is much more complicated than what these tables would lead us to
expect. Starting with tri-Hamiltonian structure the individual
entries in each one of these tables will need to be table by
itself because there are inequivalent Hamiltonian operators that
yield the same equations of motion with the same Hamiltonian
function. We shall discuss this interesting situation in a future
publication on the Chaplygin-Born-Infeld equation.

\begin{table}
\begin{tabular}{c||c|c|c}
equations of motion   & $u_t = J_2 \delta H_0 $ & $u_t = J_1
\delta H_1$ & $
 J_2 J_1^{-1} u_t = - \delta_\phi H_2 $ \\ \hline
\\[-4.5mm] local Hamiltonian op.   &  $J_2$ & $J_1$ & no
\\ \hline
\\[-4.5mm] local Lagrangian   & no & ${\cal L}_1$ & ${\cal L}_2$
\\ \hline\hline
\\[-4.5mm] modified equations & $r_t = \tilde{J}_2 \delta H_0$ & $r_t = \tilde{J}_1 \delta H_1$
& \\ \hline
\\ [-4.5mm] local Hamiltonian op.  & $\tilde{J}_2 = J_1$ & $\tilde{J}_1$ &
\\ \hline
\\ [-4.5mm] local Lagrangian  & ${\cal L}_0$ & no &
\end{tabular}
\caption{The hierarchy of local Hamiltonian structures and
Lagrangians which are local functionals of Clebsch potentials for
evolutionary system $u_t = J_1 \delta H_1 = J_2 \delta H_0$ where
$J_1$ is in the canonical Darboux form. } \label{table1a}
\end{table}

\begin{table}
\begin{tabular}{c||c|c|c}
equations of motion    &  & $u_t = J_2 \delta H_0 $ & ...  \\
\hline
\\[-4.5mm] local Hamiltonian op.   & &  $J_2$ & ...
\\ \hline
\\[-4.5mm] local Lagrangian &  & no & ...
\\ \hline\hline
\\[-4.5mm] modified equations & $\tilde{J}_1  \tilde{J}_2^{-1} r_t =
- \delta_\psi H_{-1} $& $r_t = \tilde{J}_2 \delta H_0$ & ...
\\ \hline
\\ [-4.5mm] local Hamiltonian op.  & no & $\tilde{J}_2 = J_1$ & ...
\\ \hline
\\ [-4.5mm] local Lagrangian & ${\cal L}_{-1}$ & ${\cal L}_0$ & ...
\end{tabular}
\caption{ When the Miura transformation is invertible we need to
include an additional column to the left of table 1.}
\label{table1b}
\end{table}

\section{KdV}
\label{sec-kdv}

KdV stands as the symbol of completely integrable systems. We
think we know it, but it turns out to be so rich that there is
still new information to be learned about it. We recall that KdV
\begin{equation}
 u_t + 6 u \, u_x - u_{xxx}  =0 \label{kdv}
\end{equation}
admits the Kruskal sequence of conserved Hamiltonian densities
\begin{eqnarray}
{\cal H}_{-1}^{KdV} &=& u \label{kdvcasimir1} \\
{\cal H}_{0}^{KdV} &=& \frac{1}{2} u^2 \label{kdvcasimir2} \\
{\cal H}_{1}^{KdV} &=& u^3 + \frac{1}{2} u_x^{2} \label{kdvh1} \\
{\cal H}_{2}^{KdV} &=&  \frac{5}{2} \, u^{4}  + 5 \, u \,
u_{x}^{2} + \frac{1}{2} \, u_{xx}^{\;\;2} \label{kdvh3}\\
& ... &  \nonumber
\end{eqnarray}
which are in involution with respect to Poisson brackets defined
by two Hamiltonian operators
\begin{equation}
J_1 = D, \qquad J_2 = - D^3 + 2 u D + 2 D u \label{kdvops}
\end{equation}
that form a Poisson pencil. By introducing the potential
\begin{equation}
u = \phi_x
\end{equation}
KdV can be cast into variational form with two Lagrangians
\begin{eqnarray}
{\cal L}_1^{KdV} & = & {\cal H}_{-1}^{KdV} \, \phi_t - 2 {\cal
H}_1^{KdV}  \label{lagkdv}\\
 {\cal L}_2^{KdV} & = & ( {\cal H}_{0}^{KdV} + \phi_{xx} ) \phi_t
 - 2 {\cal H}_2^{KdV} \label{pavlov}
\end{eqnarray}
which consist of the classical Lagrangian and the second
Lagrangian \cite{pavlov} respectively.\footnote{Note that there is
an error in potential term of the second Lagrangian in
\cite{nhepth}. The results that follow are stated correctly.} Here
we observe that both (\ref{lagkdv}) and (\ref{pavlov}) are
examples of our general expression (\ref{genexp}) for higher
Lagrangians.

The second application of Lenard's recursion operator to $J_1$
results in a third Hamiltonian operator which is non-local so we
cannot continue to generate higher Lagrangians. But we can use
Theorem 2 to generate new lower Lagrangians for KdV. For this
purpose we note that in both Lagrangians (\ref{lagkdv}) and
(\ref{pavlov}) we should have added the constraint $ \lambda ( u -
\phi_x )$ and written the full Lagrangian. But following the
convenient abuse of language we did not do so because it was
manifest. It is, however, necessary to write the full Lagrangian
in the case of lower Lagrangians.

According to our general construction of lower Lagrangians we
first recall the original Miura transformation
\begin{equation}
u = r^2 + r_{x} \label{miura}
\end{equation}
that brings $J_2$ to the canonical form of $J_1$ in the variable
$r$. The equation of motion for $r$ is mKdV which is different
from (\ref{kdv}) but under the substitution (\ref{miura}) we have
Miura's result
\begin{equation}
u_t + 6 u \, u_x - u_{xxx} = (D + 2 r ) \left(  r_t + 6 r^2 r_x -
r_{xxx} \right) =0 \label{miura2}
\end{equation}
so that, on shell, if mKdV is satisfied then so is KdV. Now we can
introduce the Clebsch potential for the modified field variable
\begin{equation}
r = \psi_x
\end{equation}
and write the first Lagrangian for mKdV
\begin{equation}
{\cal L}_1^{mKdV} = \psi_x \psi_t + {\cal
H}_0^{KdV}\Big|_{u=\psi_x^2+\psi_{xx} } \label{mkdvlag}
\end{equation}
in a straight-forward manner. But now enforcing the constraint in
the full Lagrangian through the Miura transformation
\begin{equation}
{\cal L}_{0}^{KdV \; full} = {\cal L}_1^{mKdV}  + \lambda ( u -
\psi_x^2 - \psi_{xx} ) \label{kdv0}
\end{equation}
we shall arrive at a new Lagrangian for KdV because the Euler
equation that comes from the first variation of this action will
be satisfied by virtue of (\ref{miura2}). Unlike (\ref{pavlov})
which is a higher Lagrangian, (\ref{kdv0}) is a lower Lagrangian
in the sense of the action of the recursion operator on the
equations of motion in the resulting Euler equation.

And the saga of KdV continues! We consider the third Hamiltonian
operator for KdV
\begin{equation}
J_3 = R^2 J_1 \label{j3kdv}
\end{equation}
which is nonlocal but the relationship between differential
substitutions and Hamiltonian structures of KdV \cite{max7}
enables us to construct another new local Lagrangian for KdV. For
this purpose we recall that the differential substitution
\begin{equation}
r = \alpha q + \frac{\varepsilon}{q} + \frac{q_x}{2 q}
\end{equation}
which transforms mKdV into twice modified KdV
\begin{equation}
q_t = \left( q_{xx} - \frac{3 q_x^2}{2 q} + \frac{6
\varepsilon^2}{q}  - 2 \alpha^2 q^3 \right)_x
\end{equation}
is a Miura transformation for (\ref{j3kdv}). This can best be seen
by the expression
\begin{equation}
J_3 = \frac{1}{2}( q^2 D + D q^2 ) - q_x D^{-1} q_x
\end{equation}
for the third non-local Hamiltonian operator for KdV in terms of
twice modified variable $q$. We recall that $J_3$ is fifth order
in $u$. We have the Miura relation
\begin{eqnarray} r_t + 6 r^2 r_x - r_{xxx} & = & \left(
\alpha - \frac{\varepsilon}{q^2} - \frac{q_x}{2 q^2} + \frac{1}{q}
D \right) \nonumber \\ && \left[ q_t - \left( q_{xx} - \frac{3
q_x^2}{2 q} + \frac{6 \varepsilon^2}{q} - 2 \alpha^2 q^3 \right)_x
\right] \label{miura3}\\ &=& 0 \nonumber
\end{eqnarray}
between modified and twice modified KdV's. Introducing the
potential for the twice modified variable $q=\chi_x$ we have
\begin{equation}
u = \Phi(\chi_x,\chi_{xx},\chi_{xxx}) \equiv
\frac{\chi_{xxx}}{\chi_x} - \frac{\chi_{xx}^{\;2}}{\chi_x^2} + 2
\alpha \chi_{xx} + \alpha^2 \chi_{x}^2 + 2 \alpha \varepsilon +
\frac{\varepsilon^{2}}{\chi_x^2} \label{uq}
\end{equation}
in terms of the original field $u$. The first Lagrangian for twice
modified KdV is simply
\begin{equation}
{\cal L}_{1}^{m_2KdV} = \chi_{x} \, \chi_{t} + {\cal H}_{-1}^{KdV}
\Big|_{ u = \Phi(\chi_{x}, \chi_{xx},\chi_{xxx} ) }
\label{mmkdvlag}
\end{equation}
and therefore the second lower Lagrangian for KdV is given by
\begin{equation}
{\cal L}_{-1}^{KdV \; full} = {\cal L}_1^{m_2KdV}  + \lambda [ u -
\Phi(\chi_{x}, \chi_{xx}, \chi_{xxx} ) ] \label{kdvlagm2}
\end{equation}
which provides another illustration of (\ref{reallag2}). This
process can be continued.

We note that an alternative to the Clebsch potential for KdV is
the Schwartzian which was pointed out by Schiff \cite{sch}. We
shall postpone consideration of Schwartzian potentials to future
work.

\section{Polytropic gas dynamics}
\label{sec-gas}

The simplest examples for applying our construction of
multi-Lagrangians consist of quasi-linear second order hyperbolic
equations that Dubrovin and Novikov \cite{dn} have called
equations of hydrodynamic type. The distinguished example in this
set consists of the Eulerian equations of polytropic gas dynamics
in $1+1$ dimensions
\begin{eqnarray}
\rho _{t} + u \, \rho _{x} + \rho \, u_{x} &=&0 \label{GD1}  \\
u_{t}+ u \, u_{x}+ \rho^{\gamma-2} \rho _{x} &=&0  \nonumber
\end{eqnarray}
and in particular for $\gamma=-1$ we have the case of Chaplygin
gas, or Born-Infeld equation that was recently shown to have a
string theory antecedent \cite{jackiw2}. This system can be cast
into quadri-Hamiltonian form \cite{gn1}. For the
Chaplygin-Born-Infeld case the complete Hamiltonian structure can
be found in \cite{annov} and its symmetries were given in
\cite{hor}. In the following we shall use the labelling $u^1 =
\rho$ and $u^2=u$.

First we have three local Hamiltonian structures of first order \cite{n1}
\begin{equation}
 J_1 = \left( \begin{array}{cc} 0 & D \\ D & 0 \end{array} \right)
= \sigma^1 D,
\label{j1}
\end{equation}
\begin{equation}
 J_2 =  \left( \begin{array}{cc}  \rho \, D  + D \, \rho &
            (\gamma-2)\,  D \, u + u \, D \\
                  D \, u + (\gamma-2)\,  u \, D   &
 \rho^{\gamma-2} D + D \, \rho^{\gamma-2}  \end{array} \right),
\label{j2}
\end{equation}
\begin{equation}
 J_3 = \left( \begin{array}{cc} u \, \rho \, D  + D \, u \, \rho &
\begin{array}{c}
D \left[ \frac{1}{2} (\gamma-2) u^2 +  \frac{1}{\gamma-1}
\rho^{\gamma-1} \right] \\ + \left[ \frac{1}{2} u^2 +
\frac{1}{\gamma-1} \rho^{\gamma-1}
\right] D \end{array} \\
\begin{array}{c}
D \left[ \frac{1}{2} u^2 +  \frac{1}{\gamma-1} \rho^{\gamma-1}
\right]  \\ + \left[
  \frac{1}{2} (\gamma-2) u^2 +  \frac{1}{\gamma-1} \rho^{\gamma-1}
\right]   D  \end{array} &
u \, \rho^{\gamma-2} \, D + D  \, u \, \rho^{\gamma-2}  \end{array} \right)
\label{j3}
\end{equation}
which form a Poisson pencil ${\cal J} = J_1 + c_1 J_2 + c_2 J_3$
with $c_1, c_2$ constants, {\it i.e.} these Hamiltonian operators
are compatible. In eq.(\ref{j1}) $\sigma^1$ is the Pauli matrix
and this is the canonical Darboux form of first order Hamiltonian
operators. The equations of polytropic gas dynamics admit two
infinite hierarchies of conserved Hamiltonians which are in
involution with respect to Poisson brackets defined by all three
of these Hamiltonian operators. In the first set, which is called
Eulerian \cite{gn1}, the Hamiltonian densities are given by
\begin{eqnarray}
{\cal H}_{-1}^E &=& \rho \label{casimir1} \\
{\cal H}_{0}^E &=& u\, \rho \label{momentum} \\
{\cal H}_{1}^E &=& \frac{1}{2} u^{2} \rho
  +\frac{1}{\gamma (\gamma -1)} \rho^{\gamma} \label{hamiltonian} \\
{\cal H}_{2}^E &=& \frac{1}{6} u^{3} \rho
+ \frac{1}{\gamma (\gamma-1)} u\, \rho^{\gamma} \label{h5}\\
{\cal H}_{3}^E &=& \frac{1}{24} u^{4} \rho
+\frac{1}{2 \gamma (\gamma -1)} u^{2} \rho^{\gamma }
+\frac{1}{2 \gamma (\gamma -1)^{2} (2\gamma -1)}
\rho^{2\gamma -1} \label{h6}\\
& ... &  \nonumber
\end{eqnarray}
where (\ref{momentum}) is the momentum, (\ref{hamiltonian})  is
the familiar Hamiltonian function, the Casimir is in
(\ref{casimir1}) and the rest consist of higher Hamiltonians.
Therefore, the Euler series is the distinguished one in the
terminology of section \ref{sec-main}. The second series
\begin{eqnarray}
{\cal H}_{-1}^L &=& u \label{casimir2} \\
{\cal H}_{0}^L &=& \frac{1}{2} (\gamma-2) u^{2}
  +\frac{1}{\gamma -1} \rho^{\gamma-1} \label{h1l} \\
{\cal H}_{1}^L &=& \frac{1}{6} (\gamma-2) u^{3}
+ \frac{1}{\gamma-1} u\, \rho ^{\gamma-1} \label{h2l}\\
{\cal H}_{2}^L &=& \frac{1}{24} (\gamma-2) u^{4} +\frac{1}{4
(\gamma -1)} u^2  \rho^{\gamma -1}
+\frac{1}{2 (\gamma -1)^2 (2 \gamma - 3) } \rho^{2 (\gamma -1)} \label{h3l}\\
& ... &  \nonumber
\end{eqnarray}
is the Lagrangian series which starts with the Casimir
(\ref{casimir2}). Note that for $\gamma=2$ this series is no
longer polynomial as logarithms will enter and the same remark
holds for integer and half-integer values of $\gamma$ in both
series.

Finally, we note that the recursion operator $R_{2}^{\;1} = J_2
J_1^{-1}$ can be used to write infinitely many Hamiltonian
operators by letting it to act $n$ times on $ J_1 $. However, in
general none of these operators will be local. In particular we
note that
\begin{equation}
R_{3}^{\;1} = J_3 \, (J_1)^{-1} \ne ( R_{2}^{\;1})^2, \qquad J_3
\ne J_2 J_1^{-1} J_2
\end{equation}
except in the case of shallow water waves where $\gamma=2$ which
admits extension to integrable dispersive equations.

Next, there is a third order Hamiltonian operator \cite{on} which
was obtained from Sheftel's remarkable recursion operator
\cite{sheftel}
\begin{equation}
J_4 = D U_x^{-1} \, D U_x^{-1} \, \sigma^1 D
\label{sheftel}
\end{equation}
where
\begin{equation}
U = \left( \begin{array}{cc} u &  \rho \\
    \frac{1}{\gamma-2}  \rho^{\gamma-2}  & u \end{array} \right)
\label{sheftelu}
\end{equation}
which is only compatible with $J_0$. Higher conserved Hamiltonians
start with the density \cite{sheftel}, \cite{verosky}
\begin{equation}
\hat {\cal H}^{SV(E)}_{-1} = \frac{\rho_x}{ u_x^{\;2} -
\rho^{\gamma - 3} \rho_x^{\;2}} \label{verosky}
\end{equation}
which is part of the Eulerian series. There is also a Lagrangian
series starting with
\begin{equation}
\hat {\cal H}^{SV(L)}_{-1} = - \frac{u_x}{ u_x^{\;2} -
\rho^{\gamma - 3} \rho_x^{\;2}} \label{verosky2}
\end{equation}
and both form new infinite hierarchies of conservation laws.

   We will be interested in the Lagrangian formulation of the equations
of polytropic gas dynamics (\ref{GD1}) that correspond to all these
Hamiltonian structures. Introducing the Clebsch potentials \cite{n3}
\begin{equation}
u=\varphi_{x},  \qquad \rho =\psi_{x} \label{potentials}
\end{equation}
we have the first Lagrangian representation for this system
\begin{equation}
{\cal L}_1^{\gamma} = {\cal H}_{-1}^L \psi_{t} + {\cal H}_{-1}^E
\varphi_{t} - 2 {\cal H}_{1}^E (\varphi _{x},\psi_{x})
\label{lag1}
\end{equation}
but using the recursion operators $J_2 \, J_1^{-1}$ and $J_3 \,
J_1^{-1}$ we find two further Lagrangians
\begin{eqnarray}
{\cal L}_{2}^{\gamma} & = & {\cal H}_{0}^L \psi _{t} + {\cal
H}_{0}^E \varphi_{t} - 2 {\cal H}_{2}^E (\varphi _{x},\psi_{x})
\label{lag2} \\
{\cal L}_{3}^{\gamma} & = & {\cal H}_{1}^L \psi _{t} + {\cal
H}_{1}^E \varphi_{t} - 2 {\cal H}_{3}^E (\varphi _{x},\psi_{x})
\label{lag3}
\end{eqnarray}
which are local functionals of the Clebsch potentials. The
Lagrangian obtained through the action of the recursion operator
$J_4 \, J_1^{-1}$ is the most interesting one. Because $J_4$ is a
third order operator, the fourth Lagrangian
\begin{eqnarray}
{\cal L}_{4}^{\gamma} & = & {\cal H}^{SV(E)}_{-1} u_{t} + {\cal
H}^{SV(L)}_{-1} \rho_{t} - 2 {\cal H}_{-1}^E (\varphi
_{x},\psi_{x})
\label{lag4} \\
{\cal L}_{4}^{\gamma}& = &  \frac{\rho_{x} u_{t} - u_{x} \rho
_{t}} {u_{x}^{2}-\rho ^{\gamma -3}\rho_{x}^{2}}- 2 \, \rho
\label{p}
\end{eqnarray}
is {\it local in the velocity fields}. This is a general property
of bi-Hamiltonian structure with a pair of first and third order
Hamiltonian operators. Here we find a remarkable situation in that
the number of Lagrangians that we can construct by repeated
application of Sheftel's recursion operator $J_4 J_1^{-1}$ is {\it
infinite} in number. All of these Lagrangians will be {\it local}
in the original field variables $\rho$ and $u$.

Now we come to lower Lagrangians that will arise from Miura
transformations. The Miura transformations that bring the
Hamiltonian operators (\ref{j2}) and (\ref{j3}) to the Darboux
form of (\ref{j1}) are point transformations for equations of
hydrodynamic type. Dubrovin and Novikov had pointed out that first
order Hamiltonian operators for equations of hydrodynamic type are
given by
\begin{equation}
J^{ik} = g^{ik} \, D - g^{im} \, \Gamma^k_{mn} u^n_x \label{dnop}
\end{equation}
where $g_{ik}$ are the components of a Riemannian metric which is
flat by virtue of the Jacobi identities. The Miura transformation
provides manifestly flat coordinates for this metric. For example
from (\ref{j2}) we find the flat metric
\begin{equation}
d s_2^2  =  \frac{2}{4 \rho^{\gamma-1} - (\gamma-1)^2 u^2} \left[
\rho^{\gamma-2} d \rho^2 - (\gamma-1) u \, d \rho \, d u + \rho \,
          d u^2 \right] \label{metrics}
\end{equation}
and it can be verified that the Miura transformation
\begin{equation}
\rho  =  r \, p  \qquad u = \frac{1}{\gamma-1} \left( r^{\gamma-1}
+ p^{\gamma-1} \right)
\end{equation}
brings it into the manifestly flat form $ 2 d r \, d p$. In these
variables we find the first modified equations of gas dynamics
\begin{eqnarray}
r_t + \frac{\gamma}{\gamma-1} (r^{\gamma-1} + p^{\gamma-1}) r_x +
\gamma r \, p^{\gamma-2} p_x =0 \label{modgas1} \\
p_t + \gamma p \, r^{\gamma-2} r_x+ \frac{\gamma}{\gamma-1}
(r^{\gamma-1} + p^{\gamma-1}) p_x =0  \nonumber
\end{eqnarray}
and linear combinations of these equations with variable
coefficients give eqs.(\ref{GD1}) of gas dynamics. Introducing the
potentials
\begin{equation} r = \chi_x , \qquad p   = \upsilon_x
\end{equation}
we have the Lagrangian
\begin{equation}
{\cal L}_{0}^{\gamma full} =   \chi_{x} \upsilon_{t} +\upsilon_{x}
\chi_{t} - 2 {\cal H}_0^E + \lambda \left( u - \frac{
\chi_x^{\gamma-1} + \upsilon_x^{\gamma-1}}{\gamma-1} \right) +
\sigma \left( \rho - \chi_x \upsilon_x \right) \label{lag0gd}
\end{equation}
where ${\cal H}_0^E$ is the momentum (\ref{momentum}) expressed in
terms of the potentials $\chi$ and $\upsilon$. Transforming to the
first modified variables $r, p$ we get $\tilde{J}_1, \tilde{J}_2=
J_1$ and $\tilde{J}_3$ defining the tri-Hamiltonian structure of
eqs.(\ref{modgas1}). Now there is a new lower Lagrangian that we
can construct from the recursion operator $\tilde{J}_1
\tilde{J}_2^{-1}$. We find
\begin{equation}
\tilde{J}_1  = \left( \begin{array}{cc} (1-\gamma) \left[ r
p^{\gamma-2} \Delta D + D  r  p^{\gamma-2} \Delta \right] &
\begin{array}{c} \left[ (\gamma-2) r^{\gamma-1} + p^{\gamma-1}
\right] \Delta D \\ + D \left[
r^{\gamma-1} + (\gamma-2) p^{\gamma-1} \right]  \Delta  \end{array} \\
\begin{array}{c}
\left[ r^{\gamma-1} + (\gamma-2) p^{\gamma-1}  \right]  \Delta D\\
+ D \left[ (\gamma-2) r^{\gamma-1} + p^{\gamma-1} \right] \Delta
\end{array}
   & (1 - \gamma ) \left[ p r^{\gamma-2} \Delta D + D  p r^{\gamma-2}
   \Delta\right]
\end{array} \right), \label{jr}
\end{equation}
$$\Delta  \equiv \frac{1}{(\gamma-1) (r^{\gamma-1} -
p^{\gamma-1})^{2} } $$ where the labelling of the variables is in
the order $r$ and $p$. The new Lagrangian is given by
\begin{equation}
{\cal L}_{-1}^{\gamma full} =   \frac{\chi_x \upsilon_{t} -
\upsilon_{x} \chi_t}{ \chi_x^{\gamma-1} - \upsilon_x^{\gamma -1}}
- {\cal H}_{-1}^E +  \lambda \left( u - \frac{ \chi_x^{\gamma-1} +
\upsilon_x^{\gamma-1}}{\gamma-1} \right) + \sigma \left( \rho -
\chi_x \upsilon_x \right) \label{lowlaggas}
\end{equation}
where the momenta do not belong to the polynomial series of
conserved Hamiltonians. However, we can identify the lower momenta
from this Lagrangian
\begin{eqnarray} {\cal H}_{-2}^{\gamma \, \pm}
= \xi_{\pm}^{\frac{3-\gamma}{\gamma-1}} \; (
\xi_+ \xi_- )^{-1/2} , \nonumber \\[2mm]
\xi^2+ (\gamma-1) u \, \xi + \rho^{\gamma-1} = 0 \nonumber
\end{eqnarray}
where $\pm$ refers to Eulerian and Lagrangian series as well as
the roots of the quadratic equation.

We now turn to the third Hamiltonian structure (\ref{j3}) defined
by the flat metric
\begin{eqnarray}
d s_3^2 & = & - \frac{8 (\gamma-1)^2 }{[(\gamma-1)^2 u^{2} -4 \rho
^{\gamma -1}]^{2}}
\Big\{ u \rho^{\gamma-2} d \rho ^{2}  \nonumber \\
 && -  \frac{1}{2 (\gamma -1) } \left[ (\gamma-1)^2 u^{2}+  4 \rho
^{\gamma -1} \right]  d \rho \, d u + u \rho \, du^{2}
\Big\}\\
&=& 2 d q \, d w \nonumber
\end{eqnarray}
and the coordinate transformation that brings it to the manifestly
flat form is given by
\begin{eqnarray}
q & = & \left[ (\gamma-1)^2 u^2 - 4 \rho^{\gamma
-1} \right]^{\frac{\gamma -3}{2(1-\gamma)}}   \label{c1} \\
w&=& \int^z  \frac{1}{\sqrt{1+\xi^2} } \, \xi^{\frac{\gamma
-3}{1-\gamma}} \;
d \xi  \label{inte} \\
z&=& \sinh \left\{ \frac{1}{2} \ln \frac{ (\gamma -1) u + 2
\rho^{(\gamma-1)/2} }{  (\gamma -1) u - 2 \rho^{(\gamma-1)/2}}
\right\} \nonumber
\end{eqnarray}
where, in general, the last integral cannot be done in closed
form. For some specific values of $\gamma$ the integral
(\ref{inte}) is elementary as in the notable case of
Chaplygin-Born-Infeld. But this paper is devoted to the general
case of polytropic gas dynamics and we shall not consider
inverting (\ref{c1}), (\ref{inte}) to obtain $u, \rho$ as
functions of $q$ and $w$. We shall only remark that after this
inversion we can obtain two more new Lagrangians.

The Lagrangians (\ref{lag1}), (\ref{lag2}) and (\ref{lag3}) for
polytropic gas dynamics are examples illustrating the general
expression (\ref{genexp}) for higher Lagrangians. For equations of
hydrodynamic type there is no dispersion and hence ${\cal G}$
vanishes identically. We have given only two (\ref{lag0gd}),
(\ref{lowlaggas}) of the four lower Lagrangians because the
integral (\ref{inte}) must be carried out before we arrive at the
second modified equations of gas dynamics which will lead to two
further new Lagrangians. Certainly the Lagrangian (\ref{p}) which
is derived from bi-Hamiltonian structure with a first and third
order operators according to (\ref{genexp}) is the most remarkable
one because this is the first time it has been possible to write a
Lagrangian for polytropic gas dynamics that is local in the
original field variables, namely the density and velocity.
Furthermore it is only the first element in an infinite series of
such Lagrangians.

\section{Kaup-Boussinesq system}

Gas dynamics with $\gamma = 2$ governs the behavior of long waves
in shallow water. From the point of view of complete integrability
it is a remarkable case, because in this case we find several
completely integrable dispersive generalizations of
eqs.(\ref{GD1}). Most prominent among them is the well-known
Kaup-Boussinesq system \cite{kaup1}
\begin{equation}
u_{t} = \left(\frac{u^{2}}{2}+\rho \right)_x \qquad  \rho_{t} =
\left(u\rho +\varepsilon ^{2}u_{xx} \right)_x \label{kbous}
\end{equation}
which admits tri-Hamiltonian structure. The first Hamiltonian
structure is given by the Hamiltonian operator (\ref{j1}) and
\begin{equation}
J_{2}^{KBq} = \left( \begin{array}{cc} D & \frac{1}{2} \, D \, u
\\[2mm] \frac{1}{2} \, u D &  \frac{1}{2}
( \rho \, D + D \, \rho ) + \varepsilon^{2} D^{3}
\end{array} \right)
\label{j2kbq}
\end{equation}
where $D^{-1}$ denotes the principal value integral, is the second
Hamiltonian operator for the Kaup-Boussinesq system. In the limit
$\varepsilon \rightarrow 0$ this Hamiltonian operator reduces to
(\ref{j2}) with $\gamma=2$. The recursion operator is given by
\begin{equation}
R^{1 \; K Bq}_{2} =\left(
\begin{array}{cc}
\frac{1}{2}u+\frac{1}{2}u_{x} D^{-1} & 1 \\ \varepsilon ^{2}
D^{2}+\rho +\frac{1}{2}\rho _{x} D^{-1} & \frac{1}{2}u
\end{array} \right)
\end{equation}
and there is a third local Hamiltonian operator obtained by the
action of the recursion operator $J_{2}^{KBq}= (R^{1 \; K
Bq}_{2})^2 J_0$ as in the $\gamma=2$ case of gas dynamics.

The conserved Hamiltonians in the Eulerian and Lagrangian series
are
\begin{eqnarray}
{\cal H}_{-1}^{KBq} & = &  \rho \label{hkb1} \\
{\cal H}_0^{KBq} & = & u \, \rho  \label{hkb2} \\
 {\cal H}_{1}^{KBq} &=& \frac{1}{2} \left( \rho u^{2}+\rho
^{2} + \varepsilon ^{2} u \, u_{xx} \right) \label{hkb3}  \\ {\cal
H}_{2}^{KBq} &=&\frac{1}{2}\left[ \rho u^{3}+3\rho
^{2}u-\varepsilon ^{2}(4u_{x}\rho _{x}+3uu_{x}^{2} ) \right]
\label{hkb5} \\ H_{3}^{KBq}&=&\frac{1}{4}u^{4}\rho
+\frac{3}{2}u^{2}\rho ^{2}+\frac{1}{2}\rho ^{3}+\varepsilon
^{4}u_{xx}^{2} \label{hkb6} \\ && -\varepsilon
^{2}(\frac{5}{2}\rho u_{x}^{2}+4uu_{x}\rho _{x}+\rho
_{x}^{2}+\frac{3}{2}u^{2}u_{x}^{2})  \nonumber
 \\ & ... &   \nonumber
\end{eqnarray}
and the degeneracy in the $\gamma = 2$ case of gas dynamics is
repeated in its dispersive generalization. In particular, the
Lagrangian and Eulerian series coincide apart from a relabelling
\begin{eqnarray}
{\cal H}_{-2}^{KBq(E)} = & u & = {\cal H}_{-1}^{KBq(L)} \nonumber
\\ {\cal H}_{-1}^{KBq(E)} = & \rho & = {\cal H}_{0}^{KBq(L)} \label{degen}
\\ & ... & \nonumber \\ {\cal H}_{-2+n}^{KBq(E)}  & =&  {\cal H}_{-1+n}^{KBq(L)}
\nonumber
\end{eqnarray}
that is dictated by the recursion operator.

With the aid of the Clebsch potentials
\begin{equation}
u = \varphi_x, \qquad \rho  =  \psi_x \label{potKBq1}
\end{equation}
we obtain
\begin{equation}
{\cal L}_{1}^{KBq} = {\cal H}_{-1}^{KBq} \varphi_{t}  + {\cal
H}_{-2}^{KBq} \psi_{t} - 2 {\cal H}_{1}^{KBq} (\varphi _{x},\psi
_{x},\varphi _{xx},\psi _{xx},...) \label{lagKBq0}
\end{equation}
for the first Lagrangian. Using the technique we have presented in
section \ref{sec-main} we shall now construct higher Lagrangians.
These three local Hamiltonian structures enable us to construct
two new Lagrangians
\begin{equation}
{\cal L}_{2}^{KBq} =( {\cal H}_{0}^{KBq} +\varepsilon ^{2}\varphi
_{xxx})\varphi _{t}+ {\cal H}_{-1}^{KBq} \psi _{t} - 2 {\cal
H}_{2}^{KBq}(\varphi _{x},\psi _{x},\varphi _{xx},\psi _{xx},...)
\label{lagKBq1}
\end{equation}
and
\begin{eqnarray}
{\cal L}_{3}^{KBq} & = & \left[  {\cal H}_{1}^{KBq} +\varepsilon
^{2} \left(  2 \psi_{xxx}+\varphi _{xx}^{2} + \varphi _{x}\varphi
_{xxx} \right) \right] \varphi_{t} \nonumber
\\ &&+ \left(  {\cal H}_{0}^{KBq} +\varepsilon^{2}\varphi_{xxx}
\right)\psi_{t} -2 {\cal H}_{3}^{KBq}(\varphi_{x},\psi _{x},...)
\label{lagKBq2}
\end{eqnarray}
for the Kaup-Boussinesq system. The determination of ${\cal
G}_{\beta ; [i]}$ is according to eq.(\ref{genexp}) with $\beta=
2, 3$ and $[2] = [1] -1$ because of the relabelling difference
(\ref{degen}) between the Lagrangian and Eulerian series. Note
that the momentum map which is the coefficient of $\phi_t$ in
(\ref{lagKBq1}) is exactly the same as the momentum in front of
$\psi_t$ in (\ref{lagKBq2}). The reason for this goes back to the
degeneration of the Eulerian and Lagrangian series into one and
the fact that it is the momentum map that is the important element
in the general construction (\ref{genexp}). In the dispersionless
limit the Lagrangians (\ref{lagKBq0}), (\ref{lagKBq1}),
(\ref{lagKBq2}) reduce to the gas dynamics Lagrangians
(\ref{lag1}), (\ref{lag2}) and (\ref{lag3}) with $\gamma=2$.

\section{Kaup-Broer System}

There is another completely integrable dispersive version of the
$\gamma=2$ case of gas dynamics which is the Kaup-Broer system
\cite{kaup1}, \cite{broer}. The triangular invertible differential
substitution
\begin{equation}
\rho =\eta +\varepsilon u_{x} \label{kbqtokbr}
\end{equation}
transforms the Kaup-Boussinesq system (\ref{kbous}) into the
Kaup-Broer system
\begin{eqnarray}
u_{t} &= & u\, u_{x} +\eta_x +\varepsilon u_{xx}  \nonumber
\\ \eta _{t} &=& \left( \eta u \right)_x -\varepsilon \eta _{xx}
\label{kbroer}
\end{eqnarray}
which also has three local Hamiltonian structures \cite{kuper}.
For the Kaup-Broer system the conserved Hamiltonians in the
Eulerian series are given by
\begin{eqnarray}
{\cal H}_{0}^{KBr} &=&u\eta \\ {\cal H}_{1}^{KBr} &=&
\frac{1}{2}[u^{2}\eta +\eta ^{2}-2\varepsilon \eta u_{x}]\\ {\cal
H}_{2}^{KBr}&=&\frac{1}{2} [u^{3}\eta +3u\eta ^{2}+6\varepsilon
\eta uu_{x}-4\varepsilon ^{2}u_{x}\eta _{x}], \\ {\cal
H}_{3}^{KBr} &=&\frac{1}{4}u^{4}\eta +\frac{3}{2}u^{2}\eta
^{2}+\frac{1}{2} \eta ^{3}+\varepsilon (\frac{3}{2}\eta
^{2}u_{x}-u^{3}\eta _{x})
\\ && +\varepsilon ^{2}(2u^{2}\eta _{xx}-\eta u_{x}^{2}-\eta
_{x}^{2})-2\varepsilon ^{3}\eta _{x}u_{xx}  \nonumber
\end{eqnarray}
which can be obtained from (\ref{hkb2})-(\ref{hkb5}) through the
substitution (\ref{kbqtokbr}). The first Hamiltonian operator for
the Kaup-Broer system is given by (\ref{j1}) and the second
Hamiltonian operator
\begin{equation} J_1^{KBr} = \left( \begin{array}{cc}  D &
 \frac{1}{2} \, D \, u + \varepsilon D^{2} \\ \frac{1}{2} \, u\, D
-\varepsilon D^{2} & \frac{1}{2} ( \eta \, D + D \, \eta )
\end{array} \right)
\label{j2kbr}
\end{equation}
can be obtained from (\ref{j2kbq}) of the Kaup-Boussinesq system
using the substitution (\ref{kbqtokbr}).

For Kaup-Broer system we introduce the potentials
\begin{equation}
\eta = w_{x}, \qquad \psi = w + \varepsilon \varphi_{x}
\label{potKBr}
\end{equation}
and arrive at the first Lagrangian
\begin{equation}
{\cal L}_{1}^{KBr} =  {\cal H}_{-1}^{KBq} \varphi_{t} +  {\cal
H}_{-2}^{KBr} w_{t}  - 2 {\cal H}_{1}^{KBr}(w_{x},\varphi
_{x},w_{xx},\varphi _{xx},...) \label{lagKBr1}
\end{equation}
but now we can derive two further Lagrangians using the recursion
operator obtained from the Hamiltonian operators (\ref{j2kbr}) and
(\ref{j1}). Following our procedure of section \ref{sec-main} we
find the second Lagrangian
\begin{equation}
{\cal L}_{2}^{KBr} = ( {\cal H}_{0}^{KBr} -2\varepsilon
w_{xx})\varphi_{t} + {\cal H}_{-1}^{KBr} w_{t}  -2 {\cal
H}_{2}^{KBr}(w_{x},\varphi _{x},w_{xx},\varphi _{xx},...)
\label{lagKBr2}
\end{equation}
which is the same as the Lagrangian of Kaup-Boussinesq system
(\ref{lagKBq2}) subject to the differential substitution
(\ref{kbqtokbr}). Similarly we find
\begin{eqnarray}
{\cal L}_{3}^{KBr} & = & ({\cal H}_{1}^{KBr}  +2\varepsilon
^{2}w_{xxx}) \varphi _{t} \label{bk} \\ && + ( {\cal
H}_{0}^{KBr}+\varepsilon \varphi_{x}\varphi_{xx} +\varepsilon ^{2}
\varphi_{xxx})w_{t} -2 {\cal H}_{3}^{KBr}(\varphi _{x},w_{x},...)
\nonumber
\end{eqnarray}
as the third Lagrangian for the Kaup-Broer equations
(\ref{kbroer}). As in the case of Kaup-Boussinesq, these
Lagrangians reduce to $\gamma=2$ gas dynamics Lagrangians in the
dispersionless limit. In the Kaup-Broer Lagrangians we find
another example of the general formula (\ref{genexp}) for
Lagrangians.

\section{Nonlinear Shr\"odinger equation}

We shall consider the nonlinear Shr\"odinger equation in the
$2$-component real version
\begin{eqnarray}
\upsilon _{t} & = & \left[\frac{\upsilon ^{2}}{2} +\eta
+\varepsilon ^{2}(\frac{\eta _{xx}}{\eta }
-\frac{\eta_{x}^{2}}{2\eta ^{2}}) \right]_x \nonumber \\ \eta_{t}
& = & (\eta \upsilon )_{x}, \label{r2nls}
\end{eqnarray}
which is a reaction-diffusion system. Again this reduces to the
$\gamma=2$ case of gas dynamics in the dispersionless limit. This
version of NLS can be obtained by another triangular differential
substitution
\begin{equation}
u = \upsilon +\varepsilon \eta _{x}/\eta \label{kbrtonls}
\end{equation}
from the Kaup-Broer system.

NLS has the same first local Hamiltonian structure (\ref{j1}) as
in the case of Kaup-Boussinesq or Kaup-Broer systems. Once again
the second Hamiltonian operator for NLS can be found by the
transformation (\ref{kbrtonls}) from the second Hamiltonian
operator (\ref{j2kbr}) of the Kaup-Broer system. Thus for the
$2$-component real version of NLS the second Hamiltonian operator
is given by
\begin{equation} J^{NLS}_2 = \left( \begin{array}{cc}
 D + \varepsilon^{2}  \left\{   \begin{array}{c}
  \eta^{-1} \, D^3 + D^3 \,  \eta^{-1} \\
-\frac{1}{2} \left[  (\eta^{-1})_{xx} \, D + D \, (
\eta^{-1})_{xx}   \right]
\end{array} \right\} & \frac{1}{2} \,
 D \, \upsilon \\  \frac{1}{2} \, \upsilon D
&\frac{1}{2} (  \eta \, D  + D \, \eta )
 \end{array} \right)
 \label{j2nls}
\end{equation}
and the conserved Hamiltonians are
\begin{eqnarray}
{\cal H}_{-2}^{NLS} &=& \upsilon \\
{\cal H}_{-1}^{NLS} &=& \eta \\
{\cal H}_{0}^{NLS} &=& \upsilon \eta  \\
{\cal H}_{1}^{NLS} &=& \frac{1}{2} \left( \eta \upsilon^2 + \eta^2 -
 \varepsilon^{2} \frac{\eta_x^2}{\eta} \right)\\
{\cal H}_{2}^{NLS} &=& \frac{1}{2} \left[ \eta \upsilon^2 + 3
\upsilon  \eta^2 + \varepsilon^{2} \left( \upsilon_x  \eta_x - 3
\frac{\upsilon \eta_x^2}{\eta}
\right) \right] \\
{\cal H}_{3}^{NLS} &=&\frac{3}{4} \eta^2 \upsilon^2 + \frac{1}{4}
\eta^3 + \frac{1}{8} \upsilon^4 \eta +
 \varepsilon^{4} \left( \frac{\eta_{xx}^{\;\;2}}{2 \eta} - \frac{
 5 \eta_x^{\;4} }{ 24 \eta^3 } \right) \\
 & & +  \varepsilon^{2} \left( \upsilon^2 \eta_{xx}
 - \frac{5}{4} \eta_{x}^{\;2}
- \frac{3}{4} \frac{\eta_{x}^{\;2}}{\eta} \upsilon^2 - \frac{1}{2}
\upsilon_{x}^{\;2}  \eta \right) \nonumber
\\ &...& \nonumber
\end{eqnarray}
which forms an infinite sequence combining both Eulerian and
Lagrangian series according to (\ref{degen}).

In order to construct the Lagrangians for NLS we introduce the
potentials
\begin{eqnarray}
\upsilon & = & z_{x} \nonumber \\ z & = & \varphi -\varepsilon \ln
w_{x}   \label{potup}
\end{eqnarray}
and the first Lagrangian
\begin{equation}
{\cal L}^{NLS}_{1}= {\cal H}^{NLS}_{-1} z_{t}+ {\cal H}^{NLS}_{-2}
w_{t}  - 2 {\cal H}_{1}^{NLS}(w_{x},z_{x},...)
\end{equation}
is the classical result. Once again we shall use the techniques of
section \ref{sec-main} to construct higher Lagrangians with the
recursion operator obtained from (\ref{j2nls}) and (\ref{j1}). We
obtain two higher Lagrangians for NLS
\begin{equation}
{\cal L}^{NLS}_{2}= {\cal H}^{NLS}_{0} z_{t}+ \left[ {\cal
H}^{NLS}_{-1} +\varepsilon ^{2} \left(\frac{w_{xxx}}{w_{x}}
-\frac{w_{xx}^{2}}{w_{x}^{2}} \right)\right] w_{t} -2 {\cal
H}_{2}^{NLS}(w_{x},z_{x},...)
\end{equation}
and
\begin{eqnarray}
{\cal L}^{NLS}_{3}&=& \left\{{\cal H}^{NLS}_{0}+\varepsilon^{2}
\left[ z_{xxx} + \left(\frac{z_{x}w_{xx}}{w_{x}} \right)_{x}
\right] \right\} w_{t} \\
&&+  \left( {\cal H}^{NLS}_{1} + 2 \varepsilon ^{2} w_{xxx}
\right) z_{t} -2 {\cal H}_{3}^{NLS}(z_{x},w_{x},...).  \nonumber
\end{eqnarray}
that are local functionals of the potentials. Here again, in the
dispersionless limit we find the $\gamma=2$ gas dynamics
Lagrangians. The remarkable strength of the general expression
(\ref{genexp}) for new Lagrangians is manifest.

\section{Boussinesq Equation}

In order to discuss the bi-Hamiltonian structure and the
Lagrangians for the Boussinesq equation in a unified framework we
first turn to its dispersionless limit. For polytropic gas
dynamics we had
\begin{equation}
\rho _{t} = (\rho u)_{x}, \qquad u_{t} =
\left(\frac{u^{2}}{2}+\frac{\rho ^{\gamma -1}}{\gamma
-1}\right)_{x}
\end{equation}
with its first nontrivial commuting flow
\begin{equation}
\rho_{y} = u_{x}, \qquad u_{y} = \left(\frac{\rho ^{\gamma
-2}}{\gamma -2}\right)_{x}
\end{equation}
both of which reduce to a second order quasi-linear wave equation
\cite{gn1}. If we express Boussinesq equation in the form
\begin{equation}
\rho_{yy} -  \left( \frac{1}{2} \rho^2  - \varepsilon^2 \rho_{xx}
\right)_{xx} = 0 \label{boussinesq}
\end{equation}
or
\begin{equation}
\rho_{y}=u_{x}, \qquad u_{y}= \left(\frac{\rho ^{2}}{2} -
\varepsilon^2 \rho_{xx}\right)_x \label{b1}
\end{equation}
as a first order evolutionary system and compare its
dispersionless limit to polytropic gas dynamics, we find that it
corresponds to the commuting flow for $\gamma =4$. The completely
integrable dispersive equation
\begin{eqnarray}
\rho _{t} & = &  \left[\rho u-2\varepsilon ^{2}u_{xx} \right]_{x},
\label{commb} \\
u_{t}& = & \left[\frac{u^{2}}{2}+\frac{1}{3}\rho ^{3}-\frac{3}{2}
\varepsilon ^{2}(2\rho \rho _{xx}+\rho _{x}^{2})+2\varepsilon
^{4}\rho_{xxxx}\right]_{x} \nonumber
\end{eqnarray}
is the commuting flow to the Boussinesq equation.

This system admits bi-Hamiltonian structure \cite{olver} with the
Hamiltonian operators (\ref{j1}) and
\begin{equation}
J^{B}_{2} = \left( \begin{array}{cc}   \rho D + D \rho   - 8
\varepsilon^{2} D^{3} & 3 u \, D + 2 u_{x}
\\[4mm] 3 D u - 2 u_{x} &  \begin{array}{c} 8 ( \rho
^{2} D + D \rho^2 ) + 8 \varepsilon ^{4} D^{5} \\ - \varepsilon
^{2} [ 5 ( \rho \, D^{3} + D^3 \rho ) - 3 ( \rho_{xx} D + D
\rho_{xx} ) ]
\end{array}
\end{array} \right)
\end{equation}
which are compatible. The conserved Hamiltonian densities for the
Boussinesq system are given by
\begin{eqnarray}
{\cal H}_{-1}^{E} &=& \rho, \\
{\cal H}_{0}^{E} &=& \rho u, \\
{\cal H}_{1}^{E} &=&\frac{1}{4}\left[2\rho u^{2}+\frac{1}{3}\rho
^{4}+\varepsilon^{2}(6\rho \rho _{x}^{2}+4u_{x}^{2})+4\varepsilon ^{4}\rho _{xx}^{2}\right],
\nonumber \\
{\cal H}_{2}^{E} &=&\frac{1}{28}\left[\frac{14}{3}\rho
u^{3}+\frac{7}{3}\rho ^{4}u+14\varepsilon ^{2}(2uu_{x}^{2}+4\rho
^{2}\rho _{x}u_{x}+3u\rho \rho
_{x}^{2})\right. \label{bcomham} \\
&& \left. +28\varepsilon ^{4}(u\rho _{xx}^{2}+\rho
_{x}^{2}u_{xx}+4\rho \rho _{xx}u_{xx})+64\varepsilon ^{6}\rho
_{xxx}u_{xxx}\right]
\end{eqnarray}
in the Eulerian sequence and we have also
\begin{eqnarray}
{\cal H}_{-1}^{L} &=& u,  \\
{\cal H}_{0}^{L} &=& u^{2}+\frac{1}{3}\rho ^{3}+\varepsilon
^{2}\rho_{x}^{2} , \\
{\cal H}_{1}^{L} &=& \frac{1}{3}u^{3}+\frac{1}{3}\rho
^{3}u-\varepsilon ^{2}u(4\rho \rho _{xx}+3\rho
_{x}^{2})+\frac{16}{5}\varepsilon ^{4}u_{xx}\rho _{xx}, \label{boussham}\\
{\cal H}_{2}^{L} &=& \frac{2}{3}u^{4}+\frac{4}{3}\rho ^{3}u^{2}+\frac{4}{45}%
\rho ^{6}+\varepsilon ^{2}(\frac{28}{3}\rho ^{3}\rho
_{x}^{2}+4u^{2}\rho
_{x}^{2}+32\rho u\rho _{x}u_{x}+8\rho ^{2}u_{x}^{2}) \nonumber \\
&&+\varepsilon ^{4}(\frac{136}{5}\rho ^{2}\rho
_{xx}^{2}-\frac{248}{5}\rho
_{x}^{4}+\frac{128}{5}uu_{xx}\rho _{xx}+\frac{16}{5}u_{x}^{2}\rho _{xx}+%
\frac{96}{5}\rho u_{xx}^{2})  \\
&&+\varepsilon ^{6}(32\rho \rho _{xxx}^{2}-\frac{592}{15}\rho _{xx}^{3}+%
\frac{64}{5}u_{xxx}^{2})+\frac{64}{5}\varepsilon ^{8}
\rho_{xxxx}^{2} \nonumber
\end{eqnarray}
in the Lagrangian sequence. The Hamiltonian function of Boussinesq
system with the first order Hamiltonian operator in Darboux form
(\ref{j1}) is $\frac{1}{2}H_{0}^{L}$. We note that the system
(\ref{b1}) for the Boussinesq equation differs from all dispersive
integrable examples we encountered earlier in that its familiar
Hamiltonian function (\ref{boussham}) is in the Lagrangian
sequence. This is because Boussinesq equation is the family of
commuting flows to the regular gas dynamics hierarchy. The first
commuting higher flow for the Boussinesq system (\ref{commb}) has
the Hamiltonian function (\ref{bcomham}) in the Eulerian series.

By introducing potentials
\begin{equation}
 u =  \varphi_{x}, \qquad  \rho  =  \psi_{x}
\end{equation}
we can obtain two local Lagrangian densities for the Boussinesq
system. First we have the classical Lagrangian
\begin{eqnarray}
{\cal L}_{1}^{B(L)} & = & {\cal H}_{-1}^{L \; \gamma=4} \psi_{y}+
{\cal H}_{-1}^{E \; \gamma=4}
\varphi_{y}-{\cal H}_{0}^{L \; \gamma=4} \label{bol1} \\
{\cal L}_{1}^{B(E)} &=& {\cal H}_{-1}^{L \; \gamma=4} \psi _{t}+
{\cal H}_{-1}^{L \; \gamma=4} \varphi_{t}-2 {\cal H}_{1}^{E \;
\gamma=4} \label{cbol1}
\end{eqnarray}
for Boussinesq system and its first nontrivial commuting flow
(\ref{commb}). The second Lagrangians are given by
\begin{equation}
{\cal L}_{2}^{B(L)}= ( {\cal H}_{0}^{E \; \gamma=4} -4\varepsilon
^{2}\varphi _{xxx})\varphi _{y}+ [ {\cal H}_{0}^{L \; \gamma=4} -
5 \varepsilon ^{2} ( \psi _{x}\psi_{xx})_x +4\varepsilon
^{4}\psi_{xxxxx} ] \psi _{y} - {\cal H}_{1}^{L \; \gamma=4}
\end{equation}
\begin{equation}
{\cal L}_{2}^{B(E)} = ( {\cal H}_{0}^{E \; \gamma=4} -4\varepsilon
^{2}\varphi _{xxx})\varphi _{t}+ [ {\cal H}_{0}^{L \; \gamma=4} -
5 \varepsilon ^{2} ( \psi _{x}\psi_{xx})_x +4\varepsilon
^{4}\psi_{xxxxx} ] \psi _{t} - 2 {\cal H}_{2}^{E \; \gamma=4}
\end{equation}
according to the general construction of Lagrangians in
(\ref{genexp}). Here we see also that the Lagrangian for the
commuting flow is obtained by flipping the Hamiltonian functions
between the Lagrangian and Eulerian series while keeping the
momenta fixed. In section \ref{sec-gas} we had constructed
Lagrangians for gas dynamics using the Hamiltonians from the
Eulerian series in the potential part of the Lagrangian. The
general formula (\ref{genexp}) can readily be used to construct
Lagrangians for the commuting flow (\ref{b1}) by this simple flip
in the potential.

\section{Conclusion}

This is the first time it has been possible to write a Lagrangian
for polytropic gas dynamics that is local in the original field
variables, namely the density and velocity. It is a result of the
general expression (\ref{genexp}) that serves to identify
immediately multi-Lagrangians for completely integrable systems.
What is even more remarkable is that this is only the first
element in an infinite series of such local Lagrangians for
polytropic gas dynamics.

It is worth emphasizing again that the scheme we have presented in
section \ref{sec-main} is a universal one for the construction of
multi-Lagrangians appropriate to evolutionary systems. The
expressions (\ref{genexp}) and (\ref{reallag2}) for Lagrangians of
completely integrable systems has general validity. We note that
(\ref{genexp}) with $\alpha=1$ is true even in the case of
non-integrable equations, provided the equations are presented in
the form of conservation laws and the system admits one further
conserved quantity, namely the Hamiltonian. We have discussed in
detail the higher Lagrangians for the completely integrable
non-linear evolution equations of polytropic gas dynamics,
Kaup-Boussinesq, Kaup-Broer, NLS and Boussinesq equations all of
which bear out the universal applicability of (\ref{genexp}) in
the construction of higher  Lagrangians. We have also presented
the lower Lagrangians (\ref{reallag2}) fully for KdV and partially
for gas dynamics owing to the difficulty of writing the second
modified variables in closed form.

The invariance group of these multi-Lagrangians and their Noether
currents should prove to be of interest in discovering new hidden
symmetries of fluid mechanics. We did not discuss this important
issue here. Recently Jackiw and co-authors \cite{jackiw} have used
hidden symmetries in the classical Lagrangian for fluid mechanics
to construct very interesting field theory models of fluid
mechanics. Multi-Lagrangians may prove to be of interest in this
connection also.

\end{document}